\documentclass[final,5p,times,twocolumn]{elsarticle}
\usepackage[T1]{fontenc}
\usepackage[utf8]{inputenc}
\usepackage{lmodern}
\usepackage{graphicx}
\usepackage{gensymb}
\usepackage[english]{babel}
\usepackage{frontespizio}
\usepackage{newtxtext,newtxmath,amsmath}
\usepackage{multicol}
\usepackage{indentfirst}
\usepackage{amsmath}
\usepackage{mathtools}
\usepackage{float}
\usepackage{subcaption}
\usepackage[table]{xcolor}
\usepackage{xcolor}
\usepackage[switch]{lineno}
\usepackage{hyperref}

\title{New detailed characterization of the residual luminescence emitted by the GAGG:Ce scintillator crystals for the HERMES Pathfinder mission}

\author[a,g]{Giovanni Della Casa\corref{cor1}}
\ead{giovanni.dellacasa@inaf.it}
\author[b]{Nicola Zampa}
\author[b,a]{Daniela Cirrincione}
\author[a,b]{Simone Monzani}
\author[b,a]{Marco Baruzzo}
\author[c,d]{Riccardo Campana}
\author[a,e]{Diego Cauz}
\author[f,a,e]{Marco Citossi}
\author[a]{Riccardo Crupi}
\author[g]{Giuseppe Dilillo}
\author[a]{Giovanni Pauletta} 
\author[f]{Fabrizio Fiore}
\author[a,b]{Andrea Vacchi}

\cortext[cor1]{Corresponding author: giovanni.dellacasa@inaf.it}

\affiliation[a]{organization={Università degli Studi di Udine}, addressline={Via A. Palladio 8}, city={Udine}, postcode={33100}, country={Italy}}
\affiliation[b]{organization={INFN sez. Trieste}, addressline={Località Padriciano 99}, city={Trieste}, postcode={34149}, country={Italy}}
\affiliation[c]{organization={INAF-OAS}, addressline={Via P. Gobetti 93/3}, city={Bologna}, postcode={40129}, country={Italy}}
\affiliation[d]{organization={INFN sez. Bologna}, addressline={Viale C. Berti Pichat 6/2}, city={Bologna}, postcode={40127}, country={Italy}}
\affiliation[e]{organization={INFN sez. Udine}, addressline={Via delle Scienze 206}, city={Udine}, postcode={33100}, country={Italy}}
\affiliation[f]{organization={INAF-OATs}, addressline={Via G.B. Tiepolo 11}, city={Trieste}, postcode={34143}, country={Italy}}
\affiliation[g]{organization={INAF-IAPS}, addressline={Via del Fosso del Cavaliere 100}, city={Roma}, postcode={00133}, country={Italy}}

\begin{document}

\begin{abstract}
The HERMES (High Energy Rapid Modular Ensemble of Satellites) Pathfinder mission aims to develop a constellation of nanosatellites to study astronomical transient sources, such as gamma-ray bursts, in the X and soft $\gamma$ energy range, exploiting a novel inorganic scintillator. 
This study presents the results obtained describing, with an empirical model, the unusually intense and long-lasting residual emission of the GAGG:Ce scintillating crystal after irradiating it with high energy protons (70 MeV) and ultraviolet light ($\sim$ 300 nm). 
From the model so derived, the consequences of this residual luminescence for the detector performance in operational conditions has been analyzed. It was demonstrated that the current generated by the residual emission peaks at 1-2 pA, thus ascertaining the complete compatibility of this detector with the HERMES Pathfinder nanosatellites.
\end{abstract}

\maketitle

\newcommand{\eqname}{Equation}

\section{Introduction}

The HERMES\footnote{\url{https://www.hermes-sp.eu}} (High Energy Rapid Modular Ensemble of Satellites) Pathfinder mission \cite{HERMES} is a project that aims to deploy a constellation of nanosatellites in the low Earth orbit (LEO), equipped with a novel specifically designed detector, to study astronomical transient sources, such as gamma-ray bursts, in the X and  soft-$\gamma$ energy range. The first six units, expected to be launched in orbit towards the end of 2023/early 2024, are being developed in the context of the pathfinder mission HERMES-TP/SP (Technological/Scientific Pathfinder). With this initial fleet the objective is to demonstrate the potential of the HERMES Pathfinder approach for the study of gamma-ray bursts. The detector installed onboard the nanosatellites exploits the “siswich” architecture \cite{siswich}, consisting in coupling a silicon detector with a scintillation crystal. To fulfill the requirements of the mission, the choice for the silicon detector fell on the silicon drift detector (SDD), while for the scintillation crystal the best candidate is the GAGG:Ce (Gd$_3$Al$_2$Ga$_3$O$_{12}$:Ce, Cerium-doped Gadolinium Aluminium Gallium Garnet), a novel inorganic scintillator. This combination allows the wide energy band observed, from $\sim$ 2 keV to a few MeV. The SDD plays a dual role: on one side it directly detects the incoming X-rays, on the other side it absorbs the optical scintillation light emitted by the GAGG:Ce when illuminated by $\gamma$-rays.

The GAGG:Ce crystal produces an unusually intense and long-lasting (even for several days) residual emission (afterglow), following its exposure to ionizing radiation \cite{afterglow}. It is essential to characterize this phenomenon to comprehend the effects on the orbiting nanosatellite detection chain, whose scintillation crystals interact with the charged particles trapped in the Van Allen belts \cite{Van Allen}. This was studied a first time in \cite{Peppe}. The analysis of the data acquired in the first campaign drew attention to some limitations that could have affected its results: there was no precise knowledge of the temperature and its variations during the measurements, the timing of the measurements with respect to the end of an irradiation step was not very accurate, the PMT bias voltage differed somewhat from a measurement set to the next, and there were no means to take account of the scintillation light due to the decay of activated nuclei within the crystal.\\

To address these issues a new measurement campaign was organised in two distinct phases: first a set of irradiation steps using an LED light, then a second one using protons like in the previous study. In this article we briefly present the setup assembled to perform this experiment, then we describe the tests executed and the resulting characterization of the afterglow, and finally we discuss the consequences for the detection system in orbit.

\section{Experimental setup}

The experiment developed to study the residual luminescence consists in observing the light emitted by the crystal both while it is being excited and subsequently (nominal dimensions 12.1 $\times$ 6.94 $\times$ 15.0 mm$^3$). Since the flux of photons it radiates changes dramatically (a few orders of magnitude), two different kinds of sensors are needed. A silicon drift detector (SDD) is employed to measure the scintillation light during the irradiation time, while a photomultiplier tube (PMT, featuring an operational gain of 10$^6$) is well fitted to observe the residual luminescence. While the SDDs are powered throughout all the experiment, the PMT has to be switched off during the irradiation, to avoid damage from exposure to intense light. To optimally characterize the investigated phenomenon it is crucial to keep the devices under test in a dark environment and to measure, and possibly control, their temperature. The importance of this conditions follow from the sensitivity to both light and temperature of the crystal and the detectors. A specific setup was built to fulfill these requirements, including a picoammeter to measure simultaneously the currents of the SDDs and PMT (at a sampling frequency of 200 Hz), and a custom cooling system. The continuous measurements of the crystal emission by the SDD cells address two important aspects of the experiment: they provide accurate timing (the anodic currents of the SDD cells and the PMT are sampled at the same times) for switching on the PMT, and allow to compare the LED excitation to that provided by the beam simply scaling the proton flux, provided by the facility (TIFPA), by the ratio of the currents measured on the SDD cells. This information is necessary for the model developed in this work to describe the mechanism responsible of the residual emission. The apparatus is also equipped with an ultraviolet LED light, having a peak emission at 310 nm (specially selected based on the absorption spectrum of the crystal \cite{UVLED}), coupled to the GAGG:Ce scintillator via a specific hole bored on the crystal support. On the opposite side of the crystal, the crystal holder is designed to leave uncovered this face to irradiate the crystal with protons.

\section{Empirical model}

The long lasting luminescence, that characterize the GAGG:Ce scintillator, is due to the creation of metastable states (traps) in the crystal, linked to intrinsic or impurity defects in its lattice. Some of the charge carriers (electrons or holes) liberated by the ionizing radiation (photons or protons in this case) can be trapped in these sites, creating the metastable states that will decay over time, freeing the charge carriers. This can happen by different processes (e.g. to the conduction band by thermal energy absorption \cite{thermal} or to nearby recombination centers by direct or thermally assisted tunneling \cite{tunneling}). Ultimately all of the charge carriers recombine, mainly through radiative paths, giving rise to luminescence. From past experiments, and the tests performed for this study, it has been observed that the GAGG:Ce has an afterglow emission that may last up to several days, even a few weeks. \\ 

The model used to describe the afterglow behaviour is similar to the one described in \cite{Peppe}: it is well-fitted by a sum of exponential decays, each one corresponding to a different species of metastable states. The main differences from the previous model are the dependence of the lifetime of the traps on the temperature and the introduction of a maximum number of metastable states available per species. Consequently, each $i$-th species is now defined by four parameters: the mean lifetime at a fixed temperature (in this case 20 $^\circ$C), $\tau_i$, its temperature coefficient, T$_{\tau, i}$, the charge carrier capture rate, $n_i$, and the maximum number of traps available, $N_{tot, i}$.\\

The equation that describes charge trapping and release from one of these species changes from the expression reported in [5] by acquiring a new term, represented within square brackets, in the updated model:

\begin{equation} \label{capture}
\frac{dN_i(t)}{dt} = n_i \left[1-\frac{N_i(t)}
{N_{tot, i}}\right] \phi - \frac{N_i(t)}{\tau_i(t)}
\end{equation}

where $N_i$ is the number of metastable states occupied and $\phi$ is the flux of charged particles irradiating the crystal. In this equation, $\tau_i$ depends on time because of the varying temperature:

\begin{equation} \label{tau} 
\tau_i(t) = \tau_{20, i} \, e^{-[T(t)-20]/T_{\tau, i}} \text{.}
\end{equation}

The number of species is estimated through the fits performed on the data acquired both with the LED light and the accelerated protons.

\section{Characterization tests}

The data acquired by the picoammeter to study the afterglow includes also two more contributions: the dark current of the PMT and the offset of the picoammeter. While the former is mainly due to the thermoionic emission of electrons and can be described using Equation \ref{eq_dark_current}, the latter is a random parameter that changes every time the setup is modified. As a result, it is inserted in Equation \ref{eq_dark_current}, and must be recalculated for every fit performed.

\begin{equation} \label{eq_dark_current}
I(T) = I_{20} \, e^{-k [20 - T(t)]} + O
\end{equation}

\begin{figure}[h!]
\centering
\includegraphics[width = \columnwidth]{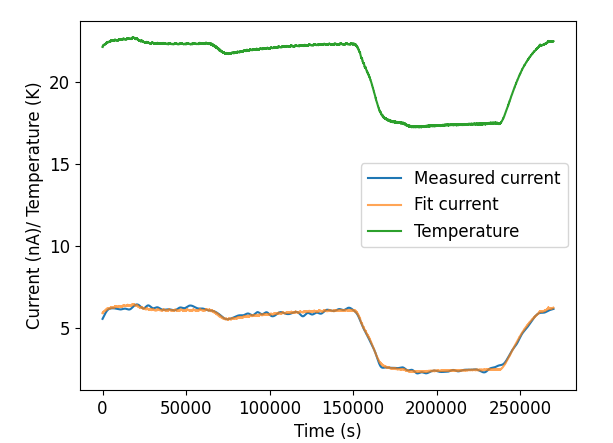}
\caption{Dark current fit superimposed on the measured data. The temperature changed in the approximate range 17 $^\circ$C to 22 $^\circ$C.}
\label{dark_current_fit}
\end{figure}

\begin{figure*}[h!]
\centering
\begin{subfigure}{0.49\textwidth}
  \centering
  \includegraphics[width=\linewidth]{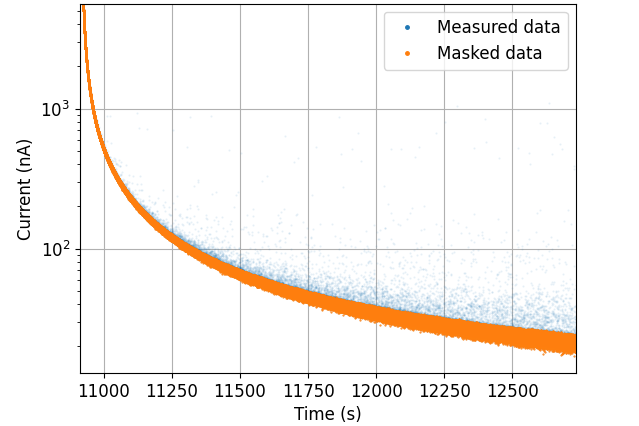}
  \caption{}
  \label{LED_single}
\end{subfigure}
\begin{subfigure}{0.49\textwidth}
  \centering
  \includegraphics[width=\linewidth]{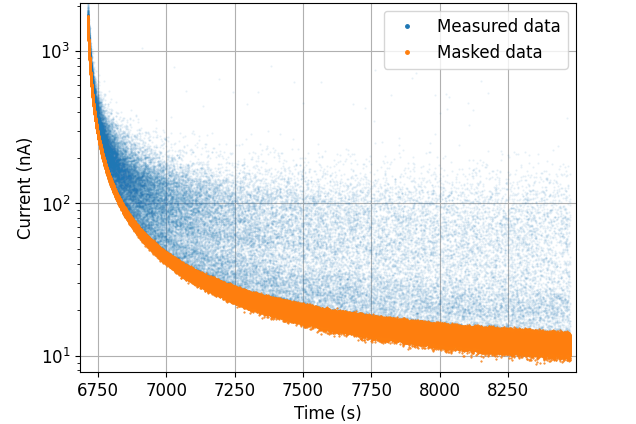}
  \caption{}
  \label{prot_single}
\end{subfigure}
\caption{Examples of the current measured by the picoammeter (in blue), after an illumination of 20 seconds at 17 $^\circ$C with the LED (left) and an irradiation of 18 seconds at 18 $^\circ$C with protons (right). In orange, the data contained within $\pm$3$\sigma$.}
\label{single_curve}
\end{figure*}

where $I$ is the current measured by the picoammeter, $I_{20}$ is the current at 20 $^\circ$C, $k$ is the coefficient that describes the temperature dependence of the dark current, and $O$ is the offset. 

For the PMT used in the following tests, the dark current parameters are those obtained with the fit of the measurements reported in \figurename~\ref{dark_current_fit}:

$$ I_{20} = 7.61 nA \;\;\; k = 0.098 K^{-1} $$

A first characterization was performed using UV LED light. This is a simpler situation, with respect to the proton beam case, that avoids the complications introduced by radiation damage. From a mathematical point of view, modelling a multiple exponential decay is a very complex problem, especially in presence of noise. The task could be simplified by leveraging the behaviour of the metastable states under varying levels of excitation. From \eqname~\ref{capture}, the number of excited states of the i-th species, $N_i$, tends to a maximum limit value when the duration of an irradiation step is much greater than $\tau_i$. While the contribution to the crystal emission from this species at the end of the irradiation depends on its duration when the exposure time is lower than $\sim\tau_i$, it becomes constant for exposures much longer than $\tau_i$. This saturation effect helps in discriminating the emission from different metastable state populations, due to their different mean lifetimes, if several irradiation steps with widely varying duration are performed. To understand the dependence of the residual luminescence from the temperature, the same set of illuminations was done at two different temperatures, about 17 and 19 $^\circ$C. Given the available time before the already planned proton irradiation campaign, the minimum time required to provide an accurate measurement of the afterglow decay following each illumination (one hour), and the need to allow measuring the very long afterglow tail to be able to capture the metastable states with large mean lifetimes, we settled to seven illumination steps at each temperature. The exposure times were set equal to 2, 5, 10, 20, 50, 100 seconds on the first day, and 500 seconds on the next day.

\begin{figure*}[h!]
\centering
\begin{subfigure}{0.48\textwidth}
  \centering
  \includegraphics[width=\linewidth]{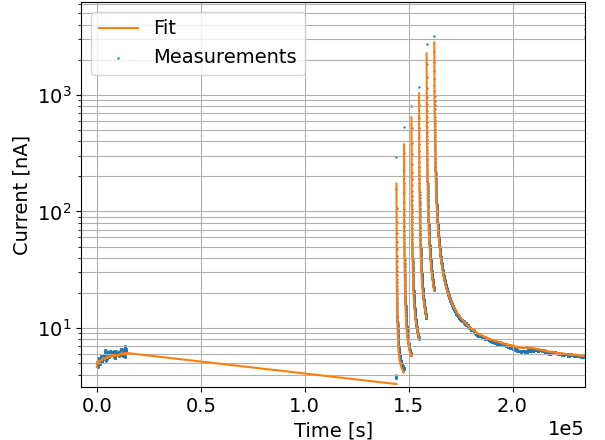}
  \label{LED_day1}
\end{subfigure}
\begin{subfigure}{0.48\textwidth}
  \centering
  \includegraphics[width=\linewidth]{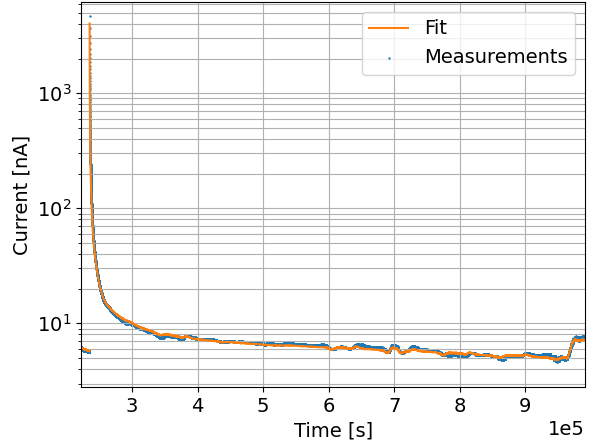}
  \label{LED_day2}
\end{subfigure}
\begin{subfigure}{0.48\textwidth}
  \centering
  \includegraphics[width=\linewidth]{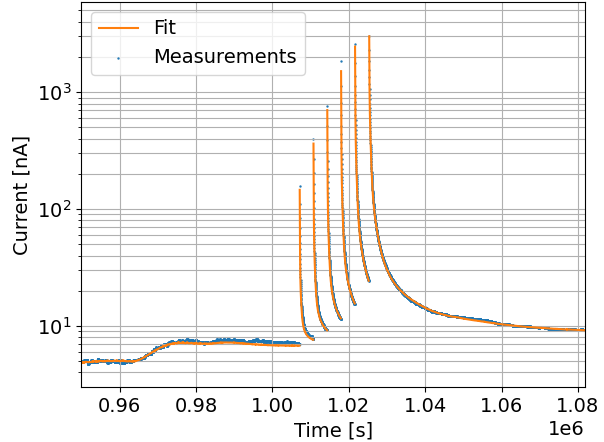}
  \label{LED_day3}
\end{subfigure}
\begin{subfigure}{0.48\textwidth}
  \centering
  \includegraphics[width=\linewidth]{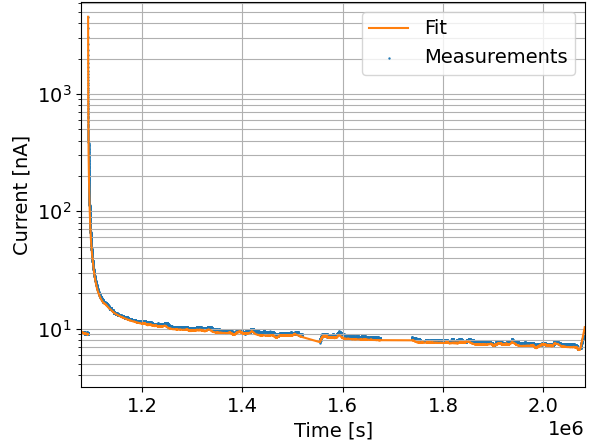}
  \label{LED_day4}
\end{subfigure}
\caption{Several plots showing the fit of the full 22 days of measurements in correspondence of the days with multiple illuminations (left column) when the low lifetime traps are more important, but also the single long illumination (right column) followed by a long observation time, when the longer lifetime traps play a major role. The plots in the upper row show measurements taken at about 17 $^\circ$C, those below at about 19 $^\circ$C. The transition between the two temperatures is shown both at the end of the upper right plot and at the begin of the lower left plot.}
\label{LED_fit}
\end{figure*}

Due to the large size of the acquired data we decided to perform a preliminary process to reduce its dimension before proceeding with the afterglow analysis. Taking as a reference \figurename~\ref{LED_single}, a spline was made from the data represented in orange, and a new set of data was extracted from this spline by selecting one point every 15 seconds for the first hour after the end of the corresponding illumination, then every minute. This procedure was repeated for each afterglow decay curve. This preliminary process is justified by the fact that the points highlighted in orange are normally distributed around the smoothing spline. The blue points in the plot show the existence of some phenomena that we were not able to identify. Since these measurements are limited to $\sim$1.5$\%$ of the data set, and their contribution is small, we neglect them in this analysis. Table \ref{LED_trap} reports the best parameters obtained by fitting the processed data with the model described before, and the corresponding curve is drawn superimposed to the data in \figurename~\ref{LED_fit}. \\

\begin{figure*}[h!]
\centering
\begin{subfigure}{0.48\textwidth}
  \centering
  \includegraphics[width=\linewidth]{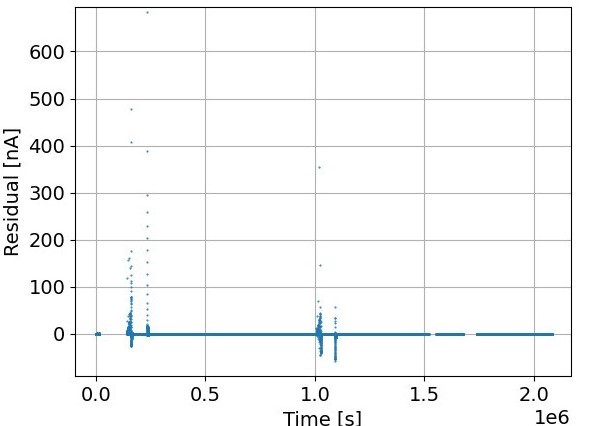}
  \label{LED_residual_norm}
\end{subfigure}
\begin{subfigure}{0.48\textwidth}
  \centering
  \includegraphics[width=\linewidth]{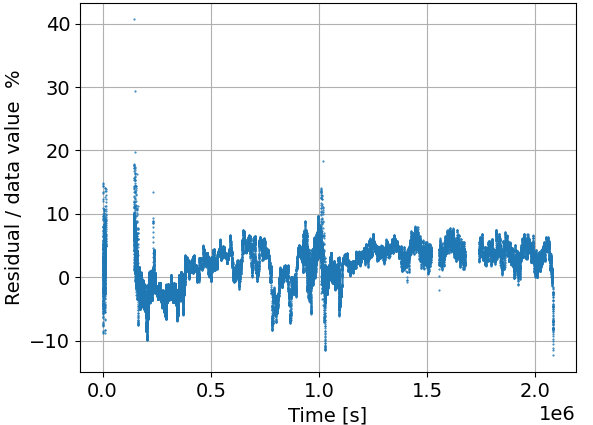}
  \label{LED_residual_per}
\end{subfigure}
\caption{The residuals obtained from the fit of the LED irradiation campaign (left) and the percentage of the residuals compared to the actual data value (right).}
\label{LED_residual}
\end{figure*}

\begin{table*}[h!]
\centering
\rowcolors{3}{white!70!black!30}{white!100!}
\renewcommand{\arraystretch}{2}
\begin{tabular}{c c c c c c c c c}
& \textbf{$\tau$} (s) & \textbf{$\sigma_{\tau}$} (s)& \textbf{$T_{\tau}$} (K) &  \textbf{$\sigma_{T_{\tau}}$} (K) & \textbf{$n$} & \textbf{$\sigma_{n}$} & \textbf{$N_{tot}$} & \textbf{$\sigma_{N_{tot}}$} \\ [0.5ex] 
\hline
\textbf{1st} & 39.95 & 0.05 & 200 & NA & 2582 & 5 & $1.201 \cdot 10^{11}$ & $1.0 \cdot 10^9$\\
\textbf{2nd} & 213.7 & 0.5 & 200 & NA & 1949 & 4 & $2.044 \cdot 10^{11}$ & $9 \cdot 10^8$ \\
\textbf{3rd} & 950.7 & 3.1 & 200 & NA & 1278 & 4 & $1.482 \cdot 10^{11}$ & $6 \cdot 10^8$ \\
\textbf{4th} & 4077 & 17 & 200 & NA & 691.9 & 9.1 & $2.468 \cdot 10^{11}$ & $7.3 \cdot 10^9$ \\
\textbf{5th} & 8174 & 374 & 200 & NA & 1700 & 54 & $1.131 \cdot 10^{10}$ & $3.6 \cdot 10^{8}$ \\
\textbf{6th} & 2.803$\cdot 10^4$ & 4.7$\cdot 10^2$ & 200 & NA & 540.9 & 18.6 & $2.925 \cdot 10^{11}$ & $4.32 \cdot 10^{10}$ \\
\textbf{7th} & 5.656$\cdot 10^5$ & 5.6$\cdot 10^3$ & 200 & NA & 3830 & 97 & $7.831 \cdot 10^{11}$ & $1.58 \cdot 10^{10}$ \\
\textbf{8th} & 2.177$\cdot 10^6$ & 8.6$\cdot 10^4$ & 1.927 & 0.072 & 3139 & 192 & $9.550 \cdot 10^{12}$ & $3.063 \cdot 10^{12}$\\
\end{tabular}
\caption{Table listing all the traps parameters for the fit of the afterglow current generated during the LED irradiation campaign in Udine.}
\label{LED_trap}
\end{table*}

The model that best matches the data is composed of eight species of metastable states. The fit gives generally a satisfying result, with a few aspects to point out. There is some difficulty in fitting the initial section of the curves following each irradiation step (more evident from \figurename~\ref{LED_residual}). This is due to limitations both in the model and in the data analysis methodology. In fact, the model we use to describe the data is concerned about metastable states with lifetimes in excess of a minimum value that is ultimately set by the timing of the data we use in the fit. As a consequence, the model is not able to fully reproduce the evolution of the luminescence signal, which features components with smaller time scales. Also with the transition between the two temperatures, it was difficult to properly estimate the temperature coefficients, especially for shorter lifetime species. For this reason, we fixed $T_{\tau, i}$ at a high value (equivalent to saying that there is no dependence of the lifetime on the temperature) for all the species except the longest one. It is interesting to notice that the transition between the two temperatures, occurring when only the states with quite long mean lifetimes are still active, is reproduced correctly by the model, confirming the relation between the temperature of the crystal and the metastable states lifetime. \\

The afterglow characterisation by means of LED excitation gives important information about the GAGG:Ce crystal, however it is not directly linked to the conditions it will experiment in the LEO radiation environment. In fact, UV light has no access to electrons whose binding energy is higher than that of the photons, while ionization has not such a limitation. Moreover, it is not able to produce two of the phenomena that occur in the interaction of energetic particles with the crystal material: activation and radiation damage. For these reasons the same crystal was subject to a second irradiation campaign at the TIFPA facility in Trento \cite{TIFPA}, using a beam of 70 MeV protons. 

\begin{table}[h!]
\rowcolors{3}{white!70!black!30}{white!100!}
\renewcommand{\arraystretch}{1.2}
\begin{center}
\begin{tabular}{c|c|c|c|c}
Day & Step & Duration (s) & Fluence & Dose (Gy) \\
\hline
Day1 & Step 1 & 2 & $3.8 \cdot 10^5$ & $2.7 \cdot 10^{-3}$ \\
& Step 2 & 7 & $1.2 \cdot 10^6$ & $8.6 \cdot 10^{-3}$ \\
& Step 3 & 18 & $3.1 \cdot 10^6$ & $2.2 \cdot 10^{-2}$ \\
& Step 4 & 30 & $5.1 \cdot 10^6$ & $3.6 \cdot 10^{-2}$ \\
& Step 5 & 470 & $8.0 \cdot 10^7$ & $5.6 \cdot 10^{-1}$ \\
Day2 & Step 1 & 30 & $5.1 \cdot 10^6$ & $3.6 \cdot 10^{-2}$ \\
& Step 2 & 48 & $8.2 \cdot 10^6$ & $5.7 \cdot 10^{-2}$ \\
& Step 3 & 67 & $1.1 \cdot 10^7$ & $7.9 \cdot 10^{-2}$ \\
& Step 4 & 82 & $1.4 \cdot 10^7$ & $9.7 \cdot 10^{-2}$ \\
& Step 5 & 800 & $1.3 \cdot 10^8$ & $9.4 \cdot 10^{-1}$ \\
Day3 & Step 1 & 30 & $5.2 \cdot 10^6$ & $3.6 \cdot 10^{-2}$ \\
& Step 2 & 130 & $2.2 \cdot 10^7$ & $1.6 \cdot 10^{-1}$ \\
& Step 3 & 220 & $3.8 \cdot 10^7$ & $2.7 \cdot 10^{-1}$ \\
& Step 4 & 1500 & $2.6 \cdot 10^8$ & $1.8$ \\
Day4 & Step 1 & 30 & $5.1 \cdot 10^6$ & $3.5 \cdot 10^{-2}$ \\
& Step 2 & 2500 & $4.2 \cdot 10^8$ & $3.0$ \\
Day5 & Step 1 & 30 & $5.2 \cdot 10^6$ & $3.6 \cdot 10^{-2}$ \\
& Step 2 & 5600 & $9.7 \cdot 10^8$ & $6.7$ \\
Day6 & Step 1 & 30 & $5.1 \cdot 10^6$ & $3.6 \cdot 10^{-2}$ \\
& Step 2 & 9600 & $1.7 \cdot 10^9$ & $11$ \\
\end{tabular}
\end{center}
\caption{Day by day schedule of all the irradiation steps performed at TIFPA, with corresponding fluence and dose.}
\label{irradiation}
\end{table}

\begin{figure*}[h!]
\centering
\begin{subfigure}{0.497\textwidth}
  \centering
  \includegraphics[width=\linewidth]{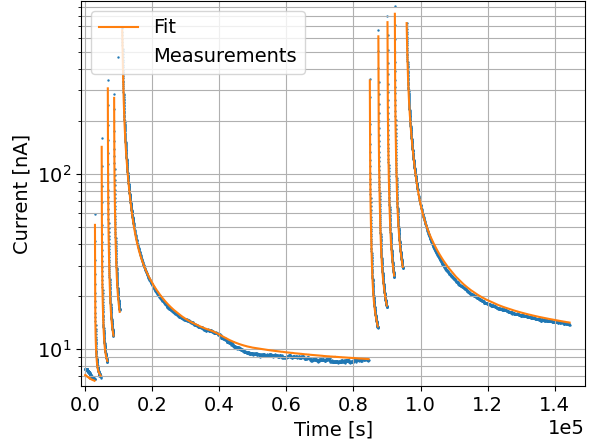}
  \label{proton_day1-2}
\end{subfigure}
\begin{subfigure}{0.497\textwidth}
  \centering
  \includegraphics[width=\linewidth]{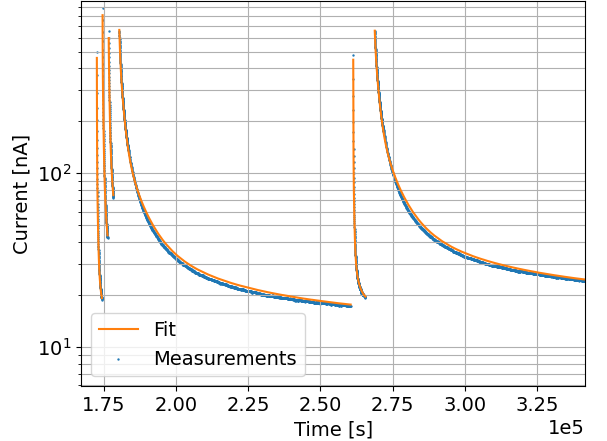}
  \label{proton_day3-4}
\end{subfigure}
\begin{subfigure}{0.51\textwidth}
  \centering
  \includegraphics[width=\linewidth]{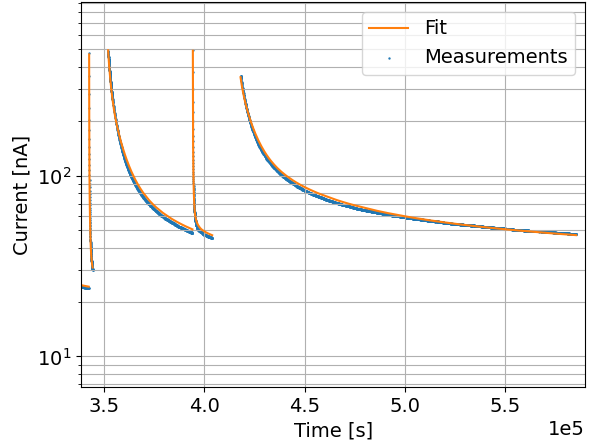}
  \label{Lproton_day5-6}
\end{subfigure}
\caption{Several plots showing the fit of the 7 days of measurements, by taking two irradiation days at a time. Following the sixth day of irradiation, there is one more day of data acquisition at rest.}
\label{proton_fit}
\end{figure*}
\begin{figure*}[h!]
\centering
\begin{subfigure}{0.497\textwidth}
  \centering
  \includegraphics[width=\linewidth]{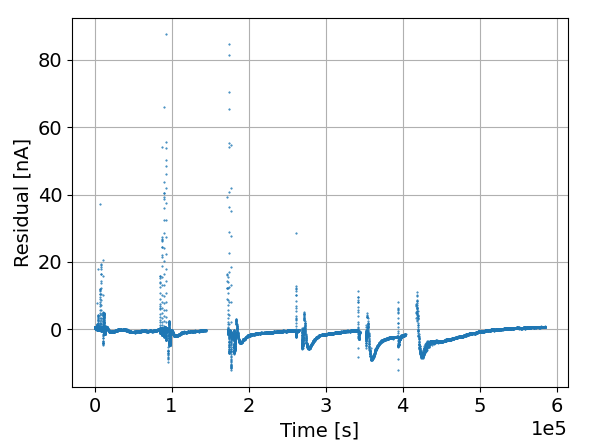}
  \label{proton_residual_norm}
\end{subfigure}
\begin{subfigure}{0.497\textwidth}
  \centering
  \includegraphics[width=\linewidth]{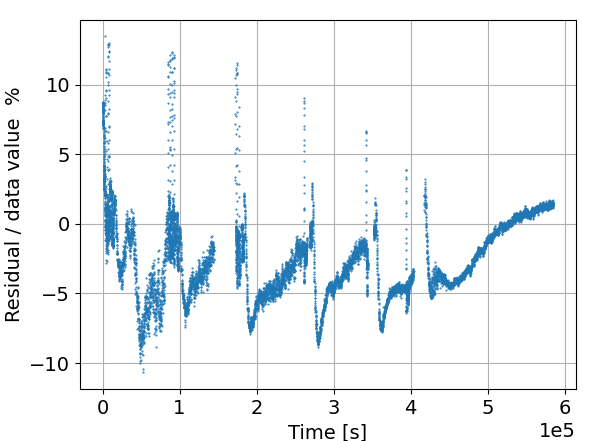}
  \label{proton_residual_per}
\end{subfigure}
\caption{The residuals obtained from the fit of the proton irradiation campaign (left) and the percentage of the residuals compared to the actual data value (right).}
\label{proton_residual}
\end{figure*}

\begin{table*}[h!]
\centering
\rowcolors{3}{white!70!black!30}{white!100!}
\renewcommand{\arraystretch}{2}
\begin{tabular}{c c c c c c c}
& \textbf{$\tau$} (s) & \textbf{$T_{\tau}$} (K) & \textbf{$n$} & \textbf{$\sigma_{n}$} & \textbf{$N_{tot}$} & \textbf{$\sigma_{N_{tot}}$} \\ [0.5ex] 
\hline
\textbf{1st} & 39.95 & 200 & 1794 & 35 & $6.177 \cdot 10^{10}$ & $2.736 \cdot 10^{10}$\\
\textbf{2nd} & 213.7 & 200 & 1222 & 1 & $1.481 \cdot 10^{13}$ & $7 \cdot 10^9$ \\
\textbf{3rd} & 950.7 & 200 & 876.3 & 1.1 & $8.784 \cdot 10^{11}$ & $1.85 \cdot 10^{10}$ \\
\textbf{4th} & 4077 & 200 & 525.7 & 1.9 & $1.156 \cdot 10^{12}$ & $4.9 \cdot 10^{10}$ \\
\textbf{5th} & 8174 & 200 & 258.6 & 2.2 & $3.177 \cdot 10^{11}$ & $9.3 \cdot 10^9$ \\
\textbf{6th} & 2.803$\cdot 10^4$ & 200 & 408.4 & 1.1 & $1.162 \cdot 10^{12}$ & $2.8 \cdot 10^{10}$ \\
\textbf{7th} & 5.656$\cdot 10^5$ & 200 & 2194 & 10 & $2.966 \cdot 10^{13}$ & $1.37 \cdot 10^{12}$ \\
\textbf{8th} & 2.177$\cdot 10^6$ & 1.927 & 7078 & 101 & $2.086 \cdot 10^{12}$ & $9.4 \cdot 10^{10}$\\
\end{tabular}
\caption{Table listing all the traps parameters, including those fixed from the LED study, for the fit of the afterglow current generated during the proton irradiation campaign at TIFPA. }
\label{proton_trap}
\end{table*}

As shown in \figurename~\ref{single_curve}, which compares two irradiation steps featuring similar duration but using UV light (left) and protons (right), the decay of the activated nuclei contribute a signal that adds to afterglow. The task of separating the two signal components can be accomplished only when the decay rate is sufficiently low to give a reasonable fraction of pure luminescence measurements. For a long-lasting irradiation this means that the first tens of seconds, or even minutes, may not be used in the study of afterglow. When a crystal is exposed to high energy radiation, some of its atoms get kicked out of their place in the lattice, producing interstitial atoms and crystal vacancies. These radiation-induced defects, called displacement damage, may give rise both to a reduced light yield, due to self-absorption by the newly formed colour centres, and to an enhancement of afterglow, via the introduction of further energy states in the band-gap. The need to study these additional phenomena led us to organize a long irradiation campaign, at TIFPA, to be performed over a continuous period of seven days, with a fixed proton flux to avoid possible complexities arising from variable dose rate effects. The crystal was subject to several irradiation steps each evening, while its emission was measured almost continuously during the whole period. For the reasons already explained, the exposure times were progressively increased from two seconds up to two hours and forty minutes, but we had to limit the measurement of afterglow decay between successive irradiation steps to half an hour in order to fulfill the schedule within the beam time allocated for our experiment on each day of the campaign (Table \ref{irradiation}). 

The analysis of the data required the addition of a new model parameter to account for a possible absorption dependent on the accumulated proton fluence. While the interactions that produce the activation of the scintillator led to the introduction of new chemical elements in the crystal, with corresponding new energy states, we expect that at the total proton fluence given to the crystal most of the metastable states that generate afterglow are still due to the same mechanisms already present at the origin. Therefore we decided to fit the data with the same mean lifetimes features ($\tau_i$, $T_{\tau, i}$) found by LED excitation, leaving free the other parameters of the metastable species ($n_i$ and $N_{tot, i}$). The afterglow curves measured at TIFPA are shown in \figurename~\ref{proton_fit}, after the preliminary processing already described, along with the model resulting from the fit. The residuals are plotted in \figurename~\ref{proton_residual}, and the parameters of the metastable states are reported in Table ~\ref{proton_trap}. \\

Like in the LED excitation case, the initial section of the curves is not reproduced very accurately by the model. The situation is better with respect to the LED characterization, mainly because we had to disregard the data affected by the decay of the activated nuclei, in which the afterglow decreases rapidly. The difficulties encountered in trying to reproduce the first portion of the curves propagate somehow to the remainders giving rise to an oscillating pattern in the residuals. Even though this is an undesired feature, we deem the result reasonable for our purposes since the residuals are relatively small (<15$\%$). 

The afterglow properties were not changed significantly by radiation damage. The reported differences in the n$_i$ and N$_{tot, i}$ parameters could be explained by the different ionization distributions between LED and proton excitations (the model considers average ionization densities), and they could also arise from the different levels of excitation attained in the two different processes (more than twice in the LED case with respect to the proton beam, according to the currents measured on the SDD cells). The analysis result is consistent with absence of radiation-induced self-absorption by color centres, with an upper limit of 0.49$\%$ at a 95$\%$ confidence level. This is in line with the results of previous measurements \cite{absorption1,absorption2,absorption3} when account is made for the much lower dose absorbed by the crystal in our experiment.

\section{In-orbit analysis}

\begin{figure}[h!]
\centering
\includegraphics[width =\columnwidth]{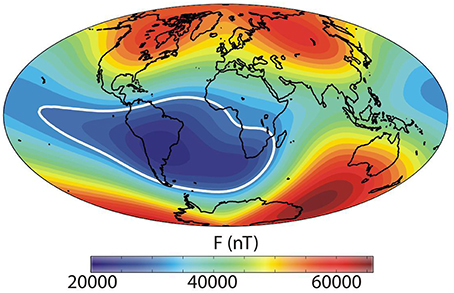}
\caption{Earth map showing the intensity of the geomagnetic field \cite{geomagnetic}.}
\label{geomagnetic}
\end{figure}

The fleet of HERMES Pathfinder nanosatellites will fly around the Earth at an altitude of 550 km on an equatorial orbit. Once in the space, the nanosatellites, and hence the crystals, will be affected by the energetic charged particles (mainly protons and electrons) trapped in the Van Allen belts \cite{Van Allen}. Flying on a low Earth orbit, the major concerns for the detectors are caused by the South Atlantic Anomaly (SAA), a region located over the south portion of the Atlantic Ocean, where the inner Van Allen belt is closer to the Earth surface due to an anomalously weak magnetic field (\figurename~\ref{geomagnetic}). \\

There have been several empirical models developed after the discovery of the Van Allen belts, a little over 60 years ago, in order to try to estimate the flux of charged particles around the Earth as a function of various parameters such as, for example, solar activity. The models that are used in general by the scientific community to simulate the orbital charged particles environments are called AP8/AE8, respectively for protons and electrons. They have been developed by NASA, using data acquired by several satellites between the sixties and the seventies. They are available on ESA’s SPace ENVironment Information System (SPENVIS). The problem with these models is that they start to be very old and they tend to underestimate the actual flux of charged particles. The former problem is quite relevant since these models are simple averages of the data taken in the two situations of maximum and minimum solar activity, that doesn't take into account changes in the last 40 years \cite{IRENE}\cite{Jakub}. In an effort to try and follow up on the models created by NASA, the IRENE (International Radiation Environment Near Earth) AE9/AP9 models were proposed by the National Reconnaissance Office (NRO) and the Air Force Research Laboratory (AFRL) \cite{IRENE}. They are based on the data acquired by 33 satellites from 1976 to 2011, so compared to the AP8/AE8 models the data used are more recent. In these models the fluxes are calculated using Monte Carlo simulations, which provides errorbars due to the uncertainties on the data and space weather changes, contrary to the AP8/AE8 models. These models too do not provide a perfect representation of the space environment, and usually overestimates the flux of charged particles \cite{Jakub}.\\

To determine whether the particle flux in orbit may generate conditions in the satellite detectors, namely high currents caused by afterglow in the crystal, that would limit the operation of the readout electronics, the AP9/AE9 models are used to create a worst case scenario since they tend to overestimate the flux of protons and electrons that could strike the devices. The front-end integrated circuits that are used by the HERMES Pathfinder nanosatellites to amplify the signal measured by the SDDs have been designed to operate with a leakage current lower than $\sim$ 500 pA, however we consider a safer limit of 100 pA to minimize non-linear behaviour when operating close to the design limit. The afterglow current generated by the decay of the metastable states is very low, so its fluctuations are not expected to trigger the detection system onboard the nanosatellites. From the point of view of the front-end electronics the current generated in the SDD by the crystal afterglow is equivalent to an additional component of leakage current, so it is important to verify that it does not reach the above mentioned limit. 

Even if the initial fleet of nanosatellites will fly on an equatorial orbit, which is more favorable since it nearly avoids the SAA, the objective of this analysis is to present a worst case scenario, that could be useful also for the following nanosatellites that will operate on different orbits. This means that instead of an equatorial orbit, we chose to analyse the case of a 10$^\circ$ orbital inclination at the nominal altitude for the HERMES Pathfinder nanosatellites, 500 km. The worst case scenario is further established by considering full efficiency for both the entrance window of the SDD (usually it is $\sim$80$\%$) and the optical coupling between the crystal and the SDD cells. 

\begin{figure}[h!]
\centering
\includegraphics[width = \columnwidth]{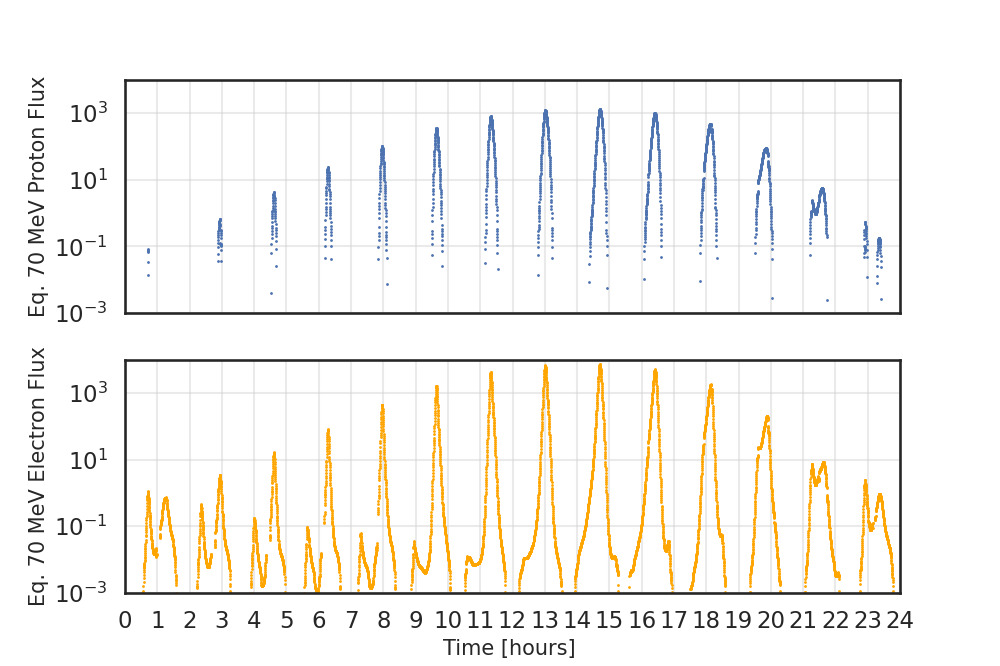}
\caption{Two plots showing 70 MeV equivalent protons and electrons flux over a day, which contains fourteen orbital revolutions around the Earth. Since one orbital revolution lasts a little more than 1.5 hours; after one day the situation goes back to the starting point. Consequently, these two plots cover all the scenarios.}
\label{AP9AE9}
\end{figure}

\begin{figure}[h!]
\centering
\includegraphics[width = \columnwidth]{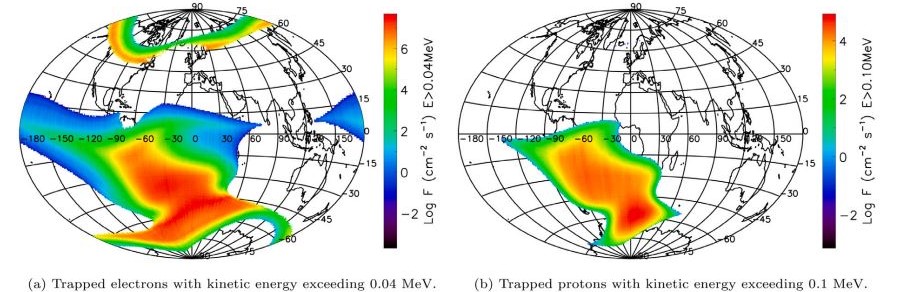}
\caption{Two world maps showing the integral flux. As expected from \figurename~\ref{geomagnetic}, almost all particle flux is concentrated on the SAA \cite{Peppe}.}
\label{IRENE}
\end{figure}

\begin{figure*}[h!]
\centering
\begin{subfigure}{\columnwidth}
  \centering
  \includegraphics[width=\linewidth]{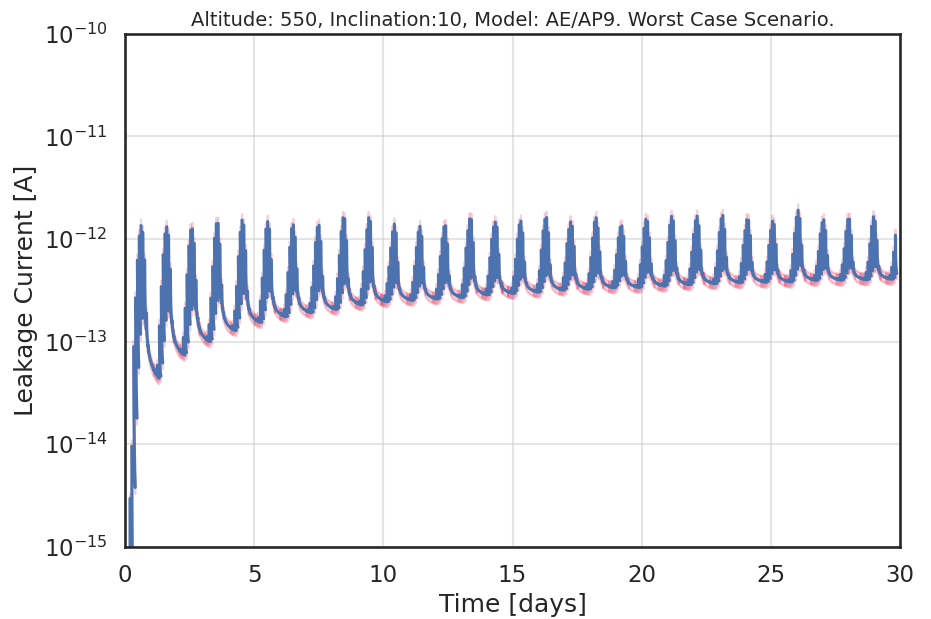}
  \label{leakage_total}
\end{subfigure}
\begin{subfigure}{\columnwidth}
  \centering
  \includegraphics[width=\linewidth]{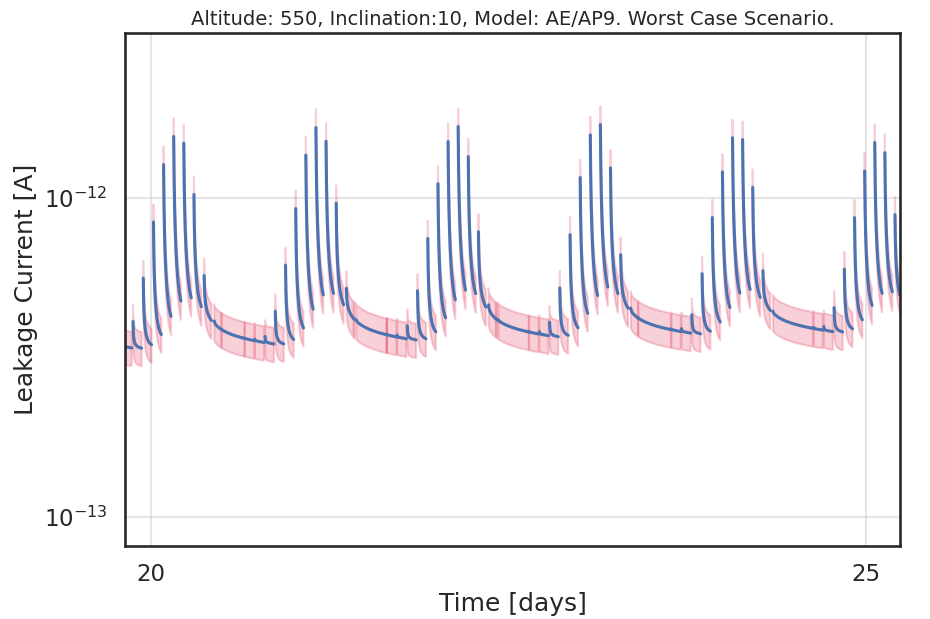}
  \label{leakage_restricted}
\end{subfigure}
\caption{Leakage current generate by the afterglow. On the left image, the entire month is shown, while on the right image only a part of it is presented, to show the effect of the single transition over the SAA. The region filled in red represent a $\pm$ 1 $\sigma$ variation on the parameters used combined with a $\pm$ 15$\%$ of the measured value, corresponding to the residual values obtained in \figurename~\ref{proton_residual}.}
\label{leakage_curr_20}
\end{figure*}

\begin{figure*}[h!]
\centering
\begin{subfigure}{\columnwidth}
  \centering
  \includegraphics[width=\linewidth]{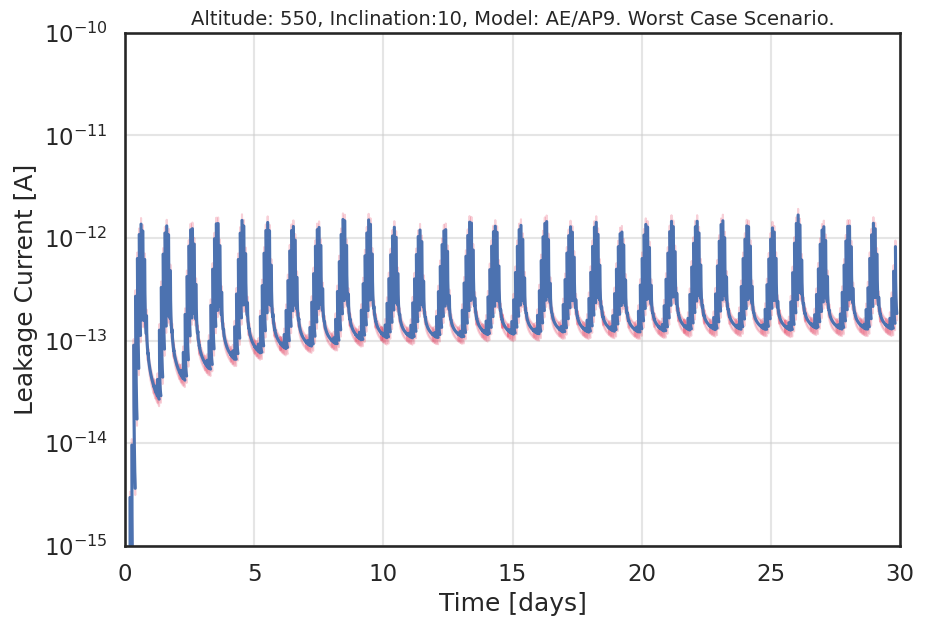}
  \label{leakage_total_5}
\end{subfigure}
\begin{subfigure}{\columnwidth}
  \centering
  \includegraphics[width=\linewidth]{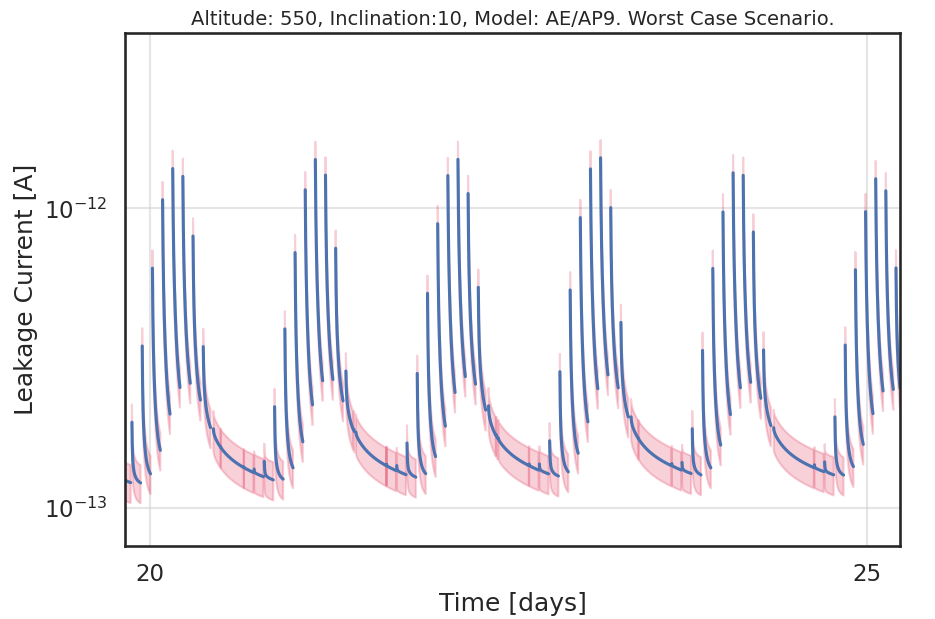}
  \label{leakage_restricted_5}
\end{subfigure}
\caption{Plots of the leakage current generate by the afterglow at 5 $^\circ$C. On the left image, the entire month is shown, while on the right image only a part of it is presented, to show the effect of the single transition in the SAA. The region filled in red represent a $\pm$ 1 $\sigma$ variation on the parameters used combined with a $\pm$ 15$\%$ of the measured value, corresponding to the residual values obtained in \figurename~\ref{proton_residual}.}
\label{leakage_curr_5}
\end{figure*}

\figurename~\ref{AP9AE9} reports, for the chosen orbit, the estimated proton and electron fluxes over an entire day, while \figurename~\ref{IRENE} illustrates the integral flux maps from which the data of the former image are calculated. The fluxes in \figurename~\ref{AP9AE9} are normalized at 70 MeV because this is approximately the maximum kinetic energy that a charged particle can have and still be stopped inside a GAGG:Ce crystal with the nominal dimensions for the HERMES Pathfinder nanosatellites. The normalization consists in multiplying the differential flux at energies lower than 70 MeV by the ratio between the proton energy and the normalization energy. The approximation for energies above the normalization value is not actually correct: these charged particles are expected to leave less than 70 MeV, since the Bragg peak, where the majority of the energy is lost, is not contained inside the crystal. However, to simplify the calculations and remain in a worst case scenario, these particles are accounted to release 70 MeV inside the scintillator. \\

All the calculation are done at a temperature of 20 $^\circ$C, in order to compare with the work done in \cite{Peppe}. However, due to the longer mean lifetime of the traps, it is important to explore the effects of the afterglow on a timescale longer than seven days. \figurename~\ref{leakage_curr_20} shows the leakage current accumulated over one month of orbital revolution around the Earth. Since the nanosatellites will be switched off during the crossing of the SAA, the corresponding data are not reported here, as well as for the minute following the exit from the SAA, that correspond to the switch on of the nanosatellite, and is also the afterglow part that is described with the largest uncertainty by the fit done in the previous sections. 

The results obtained are at least one order of magnitude larger than what was found previously. This is mainly because the metastable states with longer lifetimes could not be taken into account in the previous study, since it was impossible to observe them with the short afterglow observation time available at that time. Since these metastable states last so long before decaying, they accumulate over several orbits. Following the initial growth in the leakage current, due to the detector entering the space environment, the current tends to stabilize after some time, peaking above 1 pA. However it can be seen that the minimum value keeps growing over the month long simulation, initially being less than 0.1 picoampere, and then reaching a few tenths of a picoampere. This minimum correspond to the current of the longest traps. By the end of the first month it is still slowly increasing since the contribution of the 8th species of traps is yet to reach a plateau. The final minimum value of the afterglow current is more or less 1.2 pA, that is reached after several months. Instead the maximum seems to remain constant. \\

In this analysis, the only parameter that is not set to the actual scenario that the payload will experience is the temperature. The HERMES Pathfinder detector is expected to operate at a temperature of 5 $^\circ$C. The dependence of the mean lifetime of the metastable states on temperature is not characterized accurately in our model since we used a simplified functional description and, mainly, because of the limited temperature excursion during the measurements. To establish with more precision this temperature dependence, specific measurements should be done. An estimate of the afterglow at the operational temperature, based on our model, is nevertheless reported in \figurename~\ref{leakage_curr_5} to provide an idea of what the scenario could be at the nominal temperature. Since the only species affected by this change is the longest one, we observe a variation on the minimums of the afterglow current because the lifetime of the metastable states increases at lower temperatures, so the majority of those traps still have to decay. However, the minimum current is not expected to grow much higher than the level already reached at the end of the month. On the other hand, the maximum is still contained within a few picoamperes, slightly lower than at 20 $^\circ$C due to the lower contribution of the longest metastable states to the afterglow current, providing no risks for the nanosatellite electronics.

\section{Conclusions}

The study of the residual luminescence generated by the crystal scintillator required the development of an empirical model to explain the afterglow mechanism. The model was shown to be adequate for both LED and proton excitations. The comparison between the data acquired with the two characterization measurements evidenced the absence of significant radiation damage, at least up to the total proton fluence received by the crystal under test. The fits carried out provided the parameters necessary to simulate the residual emission generated in the crystal in operational conditions, and to verify that the current generated by the afterglow in the SDDs, acting like a leakage current, does not prevent the proper functioning of the electronics. Using a software for the simulation of the space environment (IRENE9), it was verified the fact that the afterglow current so generated reaches a peak value of a few pA (at 20 $^\circ$C), which is largely below the imposed limit value (100 pA) for the leakage current. A second estimate was performed at 5 $^\circ$C providing again comforting results.

\section*{Acknowledgments}
This study was funded by the European Union Horizon 2020 Research and Innovation Framework Programme under the grant agreement HERMES-Scientific Pathfinder n. 821896 and from ASI-INAF Accordo Attuativo HERMES Technologic Pathfinder n. 2018-10-HH.0. The research leading to these results has also received funding from the European Union’s Horizon 2020 Programme under the AHEAD2020 project grant agreement n. 871158. We kindly thank the INFN ReDSoX collaboration, the TIFPA staff and the Udine University physics workshop for their precious contributions.

Giovanni Pauletta (1945-2022) passed away during the preparation of this paper.

\end{document}